\documentclass[10pt,preprint]{aastex}
\usepackage{graphicx} 

\shortauthors{Alessandrini et al.}

\def\ltsima{$\; \buildrel < \over \sim \;$}
\def\gtsima{$\; \buildrel > \over \sim \;$}
\def\lsim{\lower.5ex\hbox{\ltsima}}
\def\gsim{\lower.5ex\hbox{\gtsima}}

\begin{document}

\title{Dynamical friction in multi-component evolving globular clusters}

\author{Emiliano Alessandrini$^1$,
Barbara Lanzoni$^1$,
Paolo Miocchi$^1$,
Luca Ciotti$^1$,
Francesco R. Ferraro$^1$}

\affil{\textsuperscript{1} Dept. of Physics and Astronomy, University of
Bologna, viale Berti Pichat, 6/2.}

\received{09/07/2014}

\begin{abstract}

We use the Chandrasekhar formalism and direct N-body simulations to
study the effect of dynamical friction on a test object only slightly
more massive than the field stars, orbiting a spherically symmetric
background of particles with a mass spectrum. The main goal is to
verify whether the dynamical friction time ($t_{\rm DF}$) develops a
non-monotonic radial-dependence that could explain the bimodality of
the Blue Straggler radial distributions observed in globular
clusters. In these systems, in fact, relaxation effects lead to a mass
and velocity radial segregation of the different mass components, so
that mass-spectrum effects on $t_{\rm DF}$ are expected to be
dependent on radius.  We find that, in spite of the presence of
different masses, $t_{\rm DF}$ is always a monotonic function of
radius, at all evolutionary times and independently of the initial
concentration of the simulated cluster. This because the radial
dependence of $t_{\rm DF}$ is largely dominated by the total mass
density profile of the background stars (which is monotonically
decreasing with radius). Hence, a progressive temporal erosion of the
BSS population at larger and larger distances from the cluster center
remains the simplest and the most likely explanation of the shape of
the observed BSS radial distributions, as suggested in previous
works. We also confirm the theoretical expectation that approximating
a multi-mass globular cluster as made of (averaged) equal-mass stars
can lead to significant overestimates of $t_{\rm DF}$ within the
half-mass radius.
\end{abstract}


\section{Introduction}
\label{intro}

Blue Straggler Stars (hereafter BSSs) are core hydrogen-burning
objects that, in the color-magnitude diagram (CMD) of globular
clusters (GCs) and other stellar systems, populate a region along the
extrapolation of the main sequence (MS) towards colors bluer than the
turn-off point \citep[see e.g.,][]{sandage53, ferraro+2009,
  mathieu+2009, monelli+2012}. This observational evidence and
direct measurements imply that the mass of BSSs is larger (in GCs,
$m_{\rm BSS} \simeq 1.2 − 1.4 M_{\odot}$; \citealp{shara97, ferraro2006a,
  lanzoni2007a, fiorentino2014}) than that of typical stars
($m_{\rm ave}\simeq 0.4 M_{\odot}$ in GCs) \footnote {Note that the typical star
  mass at the MS-TO for a 13 Gyr-old GC is $m_{\rm MS-TO}\simeq 0.8
  M_\odot$.}. In the last decades, the ratio between the number
of BSSs and that of a reference population (as red giant or horizontal
branch stars) has been derived as a function of radius for more than
$20$ Galactic GCs and three different shapes of the BSS radial
distribution have been observed \citep[see][]{ferraroNATURE}: a flat
distribution (with constant value throughout the cluster), a bimodal
trend (with a central peak, followed by a minimum and an outer rising
branch) and a unimodal behavior (with a central peak and a monotonic
decrease outward).  Some of the features of the observed BSS radial
distributions (as the radial position of the minimum in the case of
bimodal trends) have been found to be well reproduced in terms of
dynamical friction (DF) effects, even if several simplifying
assumptions were used to describe the process
\citep[see][]{mapelli2004,mapelli2006,lanzoni2007a,lanzoni2007b}.

Building on these considerations, \citet{ferraroNATURE} proposed a
comprehensive interpretative scenario identifying in the DF process
the primary cause of the whole variety of shapes of the observed BSS
radial distributions \citep[e.g.,][]{ferraro93, ferraro97,
  ferraro2004, ferraro2006b, lanzoni2007a, lanzoni2007b,
  dalessandro2008, beccari2011}.  These authors linked the three
classes of BSS radial profiles to the cluster dynamical age, thus
defining the so-called {\it{dynamical clock} for GCs: a flat, bimodal,
  and unimodal BSS radial distribution corresponds, respectively, to a
  dynamically-young, -intermediate, and -old cluster, with the
  position of the minimum of the distribution acting as a clock hand
  and marking the stage of internal dynamical evolution reached by the
  system. The idea behind this scenario is that the position of the
  minimum corresponds to the distance from the center where the DF
  time ($t_{\rm DF}$) equals the system age.  Since, in general,
  $t_{\rm DF}$ monotonically increases with radius, such a minimum is
  not yet dug in dynamically-young clusters (with a flat BSS
  distribution), it is found at intermediate radii in systems with
  intermediate dynamical age (displaying a bimodal BSS distribution),
  and it already reached the farthest outskirts in dynamically-old
  clusters (characterized by a unimodal BSS distribution).}

While the proposed scenario is simple and seems to properly account
for the observational data, one cannot exclude other more complicated,
yet realistic, possibilities, given the intrinsic complexity of DF. In
particular the present study aims at answering the following question:
\emph{could the DF time-scale develop a non-monotonic radial behavior
  at some time during the cluster evolution?} In fact, if $t_{\rm DF}$
develops a minimum at a given radius $r_{\rm min}$, the BSSs orbiting
at that distance from the center would suffer from an enhanced drag
force with respect to the other BSSs orbiting at different radii, and
a minimum in the BSS radial distribution would therefore appear at
$r_{\rm min}$ (i.e., at a place and time different from what expected
by the scenario depicted above, therefore affecting the possibility to
use the distribution of BSSs as a simple clock). Note that, while
$t_{\rm DF}$ is known to be a monotonically increasing function of
radius in most of the astrophysical problematics investigated so far,
this could not be the case for BSSs in GCs, because of three combined
conditions that could play a relevant role, namely: (i) a test
particle only slightly more massive than the field objects, (ii) a
background field made up of stars with different masses, and (iii) an
evolving and radially dependent field mass function, due to the
tendency to energy equipartition. In this paper we aim at
investigating the possibility that $t_{\rm DF}$ is not a monotonic
function of radius, by using direct N-body simulations and some
analytical results pertinent to DF in the presence of a mass spectrum.

In general terms, DF can be described as the slowing-down
of a body (the {\itshape test particle}) moving in a sea of
background particles, due to the cumulative effect of long-range
interactions (e.g., \citealt{binneytremaine, spitzer87, bertin}).
The processes underlying DF have been studied notably by
\citet{chandra43}, in the case of a test particle moving
in an infinite and homogeneous background field. In this view, DF is
interpreted as the result of the non-zero vectorial sum of the
changes in the parallel component of the relative velocity caused by
all the encounters of the test particle with the field particles (each
of them treated as an isolated two-body encounter),
as a consequence of energy conservation on each relative
orbit. 
Remarkably, in the case of an isotropic velocity distribution
function, and under the assumption of impulsive approximation, it can
be shown that (at the leading order) only field particles moving
slower than the test particle contribute to the deceleration of the
test particle.
Many authors have extended the Chandrasekhar's
framework to other more realistic physical cases, with different assumptions or
different  and more sophisticated methods
\citep{chandrasekhar+vonNeumann42, chandrasekhar+vonNeumann43,
  white48, thorne68, lee69, binney77,tremaineweinberg84,
  ostriker99, ciotti+binney2004, nipoti2008}. From
the astrophysical point of view, it is clear that 
DF plays an important role on different scales, from galaxy clusters
and their cD galaxies 
\citep[e.g.,][]{ostriker75, white76, binney77, dressler1979,
  kashlinsky87, nipoti+2004, kim2005, elzant2008}, to galaxies and their GC systems
\citep[e.g.,][]{tremaine+75, bontekoe+vanAlbada87, bertin+2003, bertin+2004,
  capuzzo_dolcetta2005, arena+2006, battisti+capuzzo2012, arca-sedda+2014},
to binary black holes at the center of early type galaxies \citep[e.g.,][]{fukushige+1992,
  vecchio+1994, milos+merritt2001}.

It is clear that in the cases mentioned above the following
assumptions are fully justified:
\begin{description}
\item {\it (i)} the mass of the test particle is much larger than the
  mass of the field ones ($m_t \gg m$).  This is a realistic
  situation, for example, when studying the sinking of GCs in galaxies
  ($m_{\rm GC} \simeq 10^5 M_\odot \gg m_{*} \simeq 1
  M_\odot$, where $m_{\rm GC}$ and $m_*$ are, respectively, the typical
  mass of GCs and that of a star in a galaxy).
\item {\it (ii)} the field particles all have the same mass.  This assumption
  becomes realistic if the previous condition is verified: when
  the test particle is much more massive than the field ones, the
  background can be safely approximated by stars with mass equal to
  their average value. 
\end{description}

Clearly the case of BSSs in GCs is a significant exception to points
{\it (i)} and {\it (ii)},
being the mass
of the test particle only slightly (2-3 times) larger than that of the field
stars, so that taking into account a mass spectrum for the background
can make significant differences.
Moreover, real GCs are composed of stars in a
relatively large range of masses (nominally from $0.1$ up to $0.8$
$M_\odot$), then also assumption {\it (ii)} is not strictly
valid. Finally, it should  be noticed that in general
\citep[e.g.,][]{bertin}, the ratio between the DF and the two-body
relaxation times is given by: 
$t_{\rm DF}/t_{\rm 2b}\propto 2m/(m_t+m)$. Hence, while in the case of massive
objects DF effects manifest on time-scales shorter than the two-body
relaxation time of the system ($t_{\rm 2b}$), for comparable masses these effects occur
on quite long times, with the tendency to mix with two-body relaxation
time effects. All these considerations make the problem of modeling
the DF action in a GC more complex. 

For the reasons above, it is not surprising that the case in which the
field particles have a mass spectrum has not been extensively
investigated in the literature.  However, the considerations presented
in \citet{ciotti2010} for a homogeneous and infinite density
background with a mass spectrum, coupled with the well known dynamical
evolution of a multi-mass GC (e.g., \citealt{spitzer87}), prompted us
to investigate in more detail the problem. In fact, the dynamical
evolution of the parent GC leads to a radially dependent
stratification of masses for the background stars, so that in practice
each radius is characterized by a different mass function, leading to
a radially dependent DF strength. Moreover, it should be noted that,
if a sort of equipartition is established in the background particles,
then some additional non trivial effect due to their velocity
distributions may take place (\citealt{ciotti2010}, and
Sect. \ref{analytic}). For example, in case of a bottom-heavy mass
spectrum, on one side the low-mass population can have a higher
density than that of high-mass stars (thus providing a proportionally
larger contribution to the total DF), but, from the other side, the
velocity of the light particles is also higher, so that their
contribution to DF is reduced, in a compensating effect.  Since DF
depends on the ``local'' conditions of the system, in terms of both
density and velocity distribution, it is not trivial to predict the
final effect on the test particle due to the interplay between the
various mass-components in the cluster. Furthermore, the dynamical
evolution of a GC produces significant changes on these distributions
in time (e.g., during and after the core-collapse stage), which, in
turn, depend on the component stellar mass, too.  Hence the problem of
modeling the DF on (the slightly heavier than average) BSSs in a
dynamically evolving (multi-mass) GCs is a quite complex task.
Therefore, what really happens to the BBS population in a GC can be
analyzed only by considering the combined effects, at each radius, of
both a radially and a time dependent mass spectrum obtained (for
example) by N-body simulations, in a self-consistent way.

Within this context, we investigate here the multifaceted nature of DF
of test particles slightly heavier than average field particles
in a dynamically evolving system with a mass spectrum, combining the
approach described in 
\citet{ciotti2010} with a set of numerical $N$-body simulations.  The
main goal is to verify whether in the presence of a multi-mass
background, the DF time-scale develops a non-monotonic radial behavior
in some stages of the system evolution,
that could provide an additional explanation to the observed variety
of BSS radial distributions.

In Section \ref{analytic} we introduce the analytic approach to the
problem of DF. In Section \ref{Nbody}, we outline the mono-mass and
multi-mass $N$-body simulations used to describe the background field
component.  The results are discussed in Section \ref{results} and
summarized in Section \ref{summary}.

\section{Analytical background}
\label{analytic}
\subsection{Mono-mass case}
\label{monomassa_analitico}
In order to introduce the case of a mass spectrum, we begin by
recalling the relevant aspects of DF in the standard case of a
background made by identical scattering masses $m$; this case is a
useful benchmark which allows to better identify the mass-spectrum
effects on DF.  The deceleration of a test particle of mass $m_t$,
moving with speed ${\bf{v}}_t$ and modulus $v_t$ in a homogeneous
background of particles of equal mass $m$, constant number density
$n$, and isotropic velocity distribution $f({\bf{v}})$, under the
effect of DF can be written as
\begin{equation}
\label{DF}
\frac{{d {{\bf{v}}_t}}}{dt} = -4\pi G^2 \ln\Lambda \ 
n \ m (m + m_t) \ \Xi(v_t) \frac {{\bf{v}}_t}{{v_t}^3},
\end{equation}
where $G$ is the gravitational constant, $\Xi(v_t)$ is the fraction of particles
slower than $v_t$ and $\ln\Lambda$ is the {\itshape
  velocity weighted Coulomb logarithm} \citep{chandra43,
  chandra_principle, binneytremaine, bertin}.
By definition, $\Xi(0) = 0$ and $\Xi(\infty) = 1$:
we recall that the sharp truncation of the function $\Xi$ for
velocities larger than $v_t$ is an approximation (not affecting our
discussion) due to the assumption of velocity isotropy of the
background and to the use of the lowest order term in the impulsive
approximation adopted to compute the two-body interactions. 
It is well known that several problems affect the direct application of
eq. (\ref{DF}) to spherical systems, as a local description of DF
\citep{bontekoe+vanAlbada87,bertin+2003,bertin+2004,arena+2006}. Nonetheless
several studies have been based on the
applications of eq. (\ref{DF}) to spherical systems.

Here we follow the same approach, for the case of interest (i.e., that
of the evolution of the BSS population of a GC), and take into account
the radial dependence of the number density by replacing $n$ with
$n(r)$.  In addition, in all our discussion we assume the test
particle to be on circular orbit in the GC potential well, so that
also $v_t$ depends on $r$:
\begin{equation}
\label{circular}
v_t(r) = \sqrt{G\frac{M(r)}{r}},
\end{equation}
where $M(r)$ is the total mass of the system enclosed within a sphere
of radius $r$.  From these assumptions, it follows that also the function $\Xi$
depends only on the radial distance from the center through $v_t(r)$ and
the local velocity distribution of the field particles.
As usual, from eq. (\ref{DF}) we can define the characteristic DF
times-scale as
\begin{equation}
\label{tdf_definizione}
t_{\rm DF} \equiv \frac{v_t}{|d {{\bf{v}}_t}/ dt |},
\end{equation}
so that in our case
\begin{equation}
\label{tdf_radiale}
t_{\rm DF}(r) = \frac{{v^3_t(r)}}{4\pi G^2 \rho(r) (m + m_t) \Xi(r) \ln\Lambda}
\end{equation}
where $\rho(r)=n(r) m$ is the mass density of the background.
This expression explicitly shows that the radial dependence of the DF
time-scale is shaped by three functions: the velocity of
the test particle, $v_t(r)$, the mass density of the field stars, $\rho(r)$,
and the relative number of stars moving slower than $v_t$ at any radius, $\Xi(r)$.
For simplicity, we neglect the possible dependence of $\Lambda$ on
radius: at the present level of approximation this seems a reasonable
assumption, due to the logarithmic nature of the associated term.
Note that the function $\Xi(r)$ is sometimes evaluated analytically by assuming a local
Maxwellian velocity distribution for the background particles, determined by the value of the
local velocity dispersion (e.g., \citealt{binneytremaine}). Here
however we avoid this additional assumption, as
we compute the function $\Xi(r)$ directly by counting the number of particles
moving slower than $v_t$ in the N-body simulation outputs (see Section
\ref{Nbody}).

\subsection{Multi-mass case}
\label{multi_analitico}
As described in the Introduction, this study is focused on the case of
a system made by the superposition of different mass components, such
as a GC with stars distributed according to a prescribed mass
spectrum.  \citet{ciotti2010} showed that in such circumstances the DF
experienced by a test particle can be significantly different from
that experienced in a single-mass background. Therefore, the natural
question arises of what happens in a GC, where the mass spectrum is
associated to the initial mass function (IMF), and the dynamical
evolution of the GC leads to a redistribution (through the tendency to
equipartition, e.g. \citealt{spitzer87}) of the density and velocity
profiles of the different mass components. It is easy to realize that
all these trends, weighted by the ratio between the mass of a BSS and
the mass of the field stars in each subcomponent of the cluster, could
lead to a quite complicate radial trend of the total DF.

In the presence of field particles with a mass spectrum, the total DF
deceleration can be split in the individual contributions due to
each single population. The $i^{th}$ population causes a deceleration
of the test particle which is again expressed by eq. (\ref{DF}); 
\begin{equation}
\label{DF_i}
\frac{{d {{\bf{v}}_t}}^{i}}{dt}(r) = -4\pi G^2 \ln\Lambda_i \ 
\rho_i(r) (m_i + m_t) \ \Xi_i(v_t) \frac {{\bf{v}}_t}{{v_t}^3},
\end{equation}
where now $\rho_i(r)=n_i(r) m_i$ is the local density of the
background component with stellar mass $m_i$.
Due to the additive nature of scattering effects in the Chandrasekhar
treatment of DF, the total deceleration is obtained from the sum of all the contributions.
In particular, from eqs. (\ref{tdf_definizione}) and (\ref{DF_i}) it follows that: 
\begin{equation}
\label{somma_armonica}
\frac{1}{t_{\rm DF}(r)} = \sum_{i = 1}^{N_{\rm pop}} \frac{1}{{t^{i}_{DF}(r)}}, 
\end{equation}
where
\begin{equation}
\label{tdf_i}
{{t^{i}_{DF}(r)}} = \frac{{v^3_t(r)}}{4\pi G^2 \rho_i(r) (m_i + m_t) \Xi_{i}(r) \ln\Lambda_i}.
\end{equation}
In eq. (\ref{somma_armonica}) we assumed that the mass spectrum of the
background stars is represented by the sum of a finite number
of components ($N_{\rm pop}$). However eq. (\ref{somma_armonica}) can be easily
written as an integral in
the case of a continuous mass spectrum \citep{ciotti2010}.
As a general comment, 
note that eq. (\ref{somma_armonica}) indicates that the local
value of the DF time is roughly determined by the smallest among the various $t^i_{DF}(r)$.

So far, the description in this Section just reflects the "standard"
approach to our problem: namely, one could construct an equilibrium
model for a GC (for example by solving numerically the Poisson equation
for one- and multi-component \citet{king66} or \citet{wilson75}
models), take the radial profile of $n_i(r)$
and the local velocity distribution and compute
eq. (\ref{tdf_i}).
Here, we follow a more realistic approach, namely we make use of
(collisional) N-body simulations of a set of mono- and
multi-mass GC models, capable to provide us with the self-consistent
radial behaviors of all the various quantities needed to
evaluate eq. (\ref{tdf_i}). This
approach, at variance with the solution of the Poisson problem, allows
us to take into account also the time evolution of the $t_{\rm DF}$ radial profile.

\section{$N$-body simulations}
\label{Nbody}
Here we describe the set-up of the N-body simulations performed in
order to determine the time evolution of the GC models hosting the BSS
population.  Since GCs are collisional systems, very accurate and
specifically designed numerical methods are required to properly
describe their time evolution.  In practice, we use N-body simulations
to obtain a self-consistent description of the phase-space density
distribution of the different components of the GC, and then we apply
the equations in Section \ref{analytic} to estimate the radial trend
of $t_{\rm DF}$.  We also perform some simulation where the background
stars are all characterized by the same mass, so that the effects of a
mass spectrum can be better appreciated through comparison.  For our
simulations we used the direct $N$-body code {\tt{NBODY6}}
\citep{aarseth2003}.  In all cases, the simulations are not meant to
describe the evolution of a GC from its formation to the present
days. They just provide a simplified ``picture'' of a current
  GC, which could be in a pre- or in a post-core collapse phase.

\subsection{Mono-mass simulations}
The mono-mass system is composed of $N = 10^4$ particles with mass
$m$.  The initial conditions (particle positions and velocities) have
been generated from a King \citeyearpar{king66} model with central
dimensionless potential $W_0 = 4$.  We followed the dynamical
evolution of the cluster up to a final time $t_f = 1000$ in N-body time
units (i.e. in units in which the total mass of the cluster is $M =
1$, $G = 1$ and the total energy is $E = -1/4$;
\citealp[see][Sect. 7.4]{aarseth2003}). A characteristic time scale
for two-body relaxation is given by 
$t_{\rm rh}(0)$, the half-mass relaxation time of the initial conditions
 \citep[e.g.,][]{spitzer87}:
\begin{equation}
\label{t_rh_equation}
t_{\rm rh}(0) = 0.138 N \sqrt{ \frac{r_h(0)^3}{G M} } \frac{1}{\ln(0.4 N)}
= 2.2 \, \frac{N}{\ln(0.4N)} \left(\frac{r_h(0)}{1
    \mathrm{pc}}\right)^{3/2} \left(\frac{1
    \mathrm{M}_\odot}{M}\right)^{1/2} [\mathrm{Myr}],
\end{equation}
where $r_h(0)$ is the half mass radius (i.e., the radius containing half
the total mass of the system) at the initial time $t=0$. In N-body
units $t_{\rm rh}(0) \simeq 210$
so that in practice we follow the GC evolution up to $t_f\simeq 4.8 t_{\rm rh}(0)$.
To improve the statistics, $20$ different sets of initial conditions
have been generated by changing only the random seed from which the
positions and velocities of the particles are extracted starting from
the distribution function. All the
realizations have then been combined at each time-step, shifting the
center of mass of each system to a common origin. This procedure
generated ``{\itshape
  supersnapshots}'' containing $N_{\rm super} = 2 \times
10^5$ particles.
The simulations have been run on a dedicated workstation, and each
simulation in the mono-mass case took approximately 3 hours (thanks
to the use of a GPU card).

\subsection{Multi-mass simulations}
\label{simulazioni_multi}
As described in Section 2.2, in order to determine the time evolution of the radial trend of $t_{\rm DF}$ for
the population of BSS in a multi-mass GC, we need the
distribution function of each stellar component of the cluster, i.e.,
the associated density profile and its velocity distribution. Of course, some
educated guess could be used to describe the radial profile of the
velocity distribution (e.g., to solve the associated Jeans equations
and use the resulting velocity dispersion in the local Maxwellian
approximation), but here we
prefer to use direct N-body simulations that allow to compute in a
self-consistent way the evolution of the structural and dynamical
properties of the different components.

The modelization of a mass-spectrum with a necessarily limited number
of particles imposes some constrain in the choice of the number of
mass bins. Of course, the larger is the number of bins, the finer is
the spectrum.  However, to avoid too little numbers of particles in
each bin as a consequence of an excessive partition among the field
particles, as well as to understand more clearly the contribution of
the various mass ranges (at the various radii) to the resulting DF, we
represent the mass spectrum as the superposition of three different
populations with a total number of stars $N_1$, $N_2$, and $N_3$. The
masses of the individual stars in each population are $m_1$, $m_2$,
and $m_3$ respectively, with $m_1$ being the smallest value,
$m_2=2m_1$ and $m_3=3m_1$.  Therefore, $N = N_1+N_2+N_3$ and
$m_1=M/(N_1+2N_2+3N_3)$, where $M$ is the total mass of the
cluster. The three populations are aimed at grossly representing the
main populations of field stars in a present-day GC, namely low-MS
stars (with mass $m_1 \simeq 0.3 M_{\odot}$), intermediate-MS objects
($m_2 \simeq 0.6 M_{\odot}$), and TO and giant stars ($m_3 \simeq 0.9
M_{\odot}$).  As in the mono-mass simulations, the total initial
number of stars is $N = 10^4$, with the three components counting $N_1
= 8500$, $N_2 = 1200$ and $N_3 = 300$ particles, respectively. This
choice is a reasonable compromise between the description of a
realistic case, and the need of a large statistical sampling for the
less numerous population ($m_3$) in order to avoid noise-dominated
results (see Sect. \ref{summary} for a discussion of the impact of
such approximation on the results obtained).

The initial positions and velocities of all particles in each of the
three groups are randomly extracted from the same \citet{king66}
distribution function, i.e., no initial mass segregation is
assumed. This choice is made both to avoid an additional degree of
freedom (an unconstrained amount of mass segregation) and because it
is justified by the flat BSS distribution observed in $\omega$ Cen,
Palomar 14 and NGC 2419 \citep[see][]{ferraroNATURE}.  The system is
also fully isolated, with no primordial populations of binaries or
multiple systems (see Sect. \ref{summary} for the discussion of these
approximations).

In order to improve the statistics, as in the mono-mass case, we
generated $20$ sets of initial conditions
by varying only the random seed and we combined all the $19$ runs
together at every extracted snapshot, thus generating supersnapshots
made of $N_{\rm super} \simeq 2 \times 10^5$ stars (this
number varies slightly during the evolution due to the loss of unbound
particles).  All the simulations have been stopped at the time $t_f
\simeq 10 t_{\rm rh}(0)$, with $t_{\rm rh}(0) = 190$ in $N$-body units.
In these multi-mass cases, the half-mass relaxation time is again
computed by using eq. (\ref{t_rh_equation}). We extracted a snapshot every $5 \times 10^{-3}
t_{\rm rh} (0)$, thus we have guaranteed a good accuracy in tracking the
cluster evolution. Three different values of the King dimensionless
potential ($W_0 = 4,6,8$) have been considered for the multi-mass
simulations, so in total we ran $60$ simulations in the multi-mass case.

\section{Results}
\label{results}
\subsection{Mono-mass case}
\label{results-mono}
In the mono-mass case, the DF time-scale of a test particle of mass
$m_t = 4 m$ ($m$ being the mass of the background stars) has been
evaluated by using eq. (\ref{tdf_radiale}).  In order to extract the
radial profiles of $v_t(r)$, $n(r)$ and $\Xi(r)$ entering
eq. (\ref{tdf_radiale}) from the simulations,
we considered equally populated radial bins (i.e., 100 concentric
spherical shells, each enclosing
$2000$ particles of a given supersnapshot).  This choice has the useful property of
maintaining constant the error bars of the number density over the
whole radial range,
thus reducing the effects of random fluctuations.  This choice also
implies a finer
sampling of the innermost regions as the time passes, because the
density progressively increases in the core of the GC.  The number density profile of
the field component, $n(r)$, is then given by the number of
particles in each radial bin divided by the volume of the shell. The
circular velocity of the test particle, $v_t(r)$, is trivially
computed by using eq. (\ref{circular}). 
Finally, the local estimate of
$\Xi(r)$ in each radial bin is obtained by normalizing the number of background
particles in the given radial bin that are slower than
$v_t(r)$, to the total number of background particle in the same bin.

\begin{figure}[htbp]
\centering
\includegraphics[height=17.0cm, width=17.0cm]{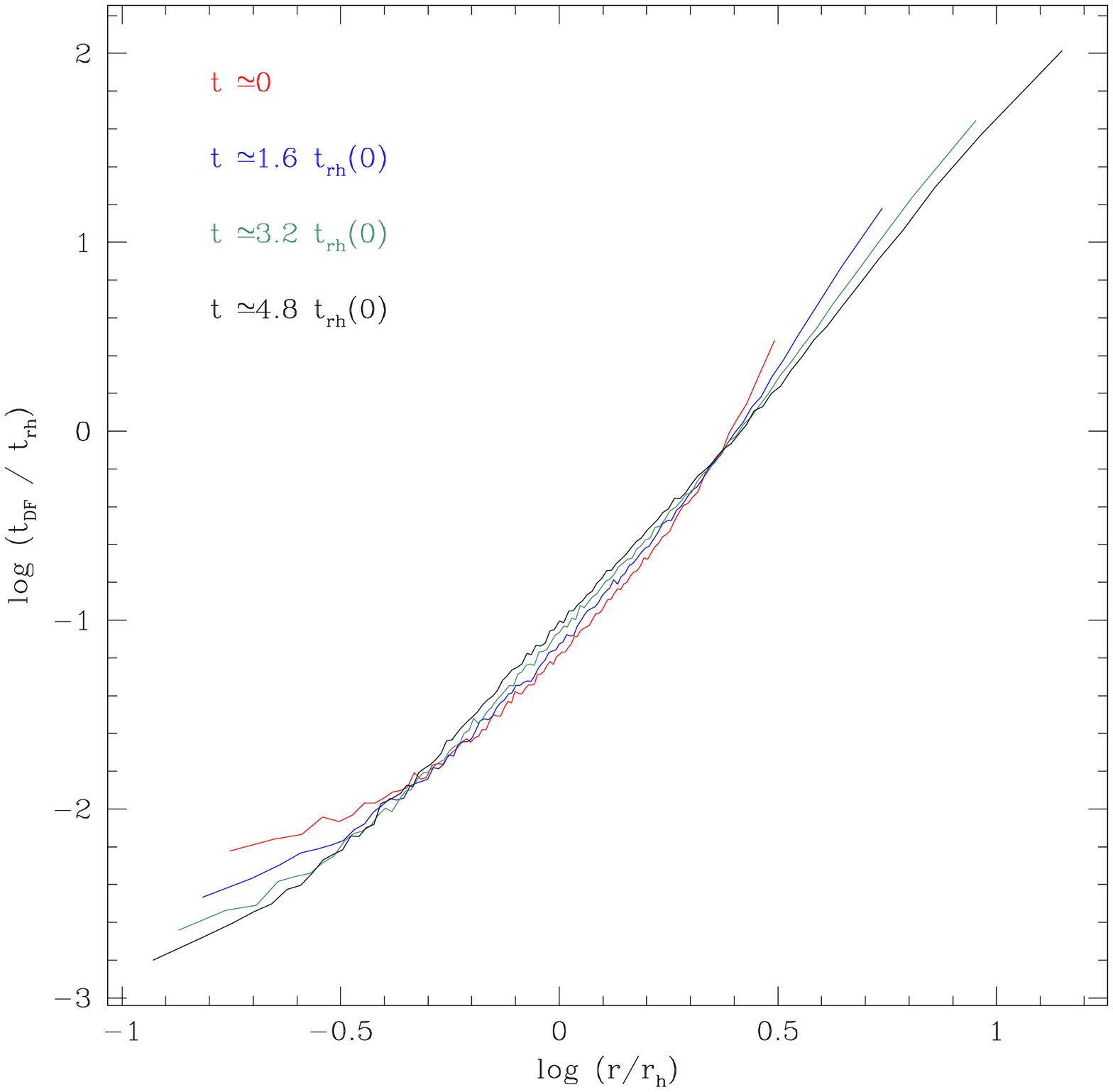}
\caption{Time evolution of the radial profile of the DF time-scale, $t_{\rm DF}$, for the
  mono-mass system with central dimensionless potential $W_0 = 4$. The
  four curves correspond to four different times of the simulation,
  from $t = 0$ to the final time $t_f = 4.8 t_{\rm rh}(0)$. As explained
  in the text, the DF
  time-scale is normalized to the instantaneous half-mass relaxation
  time, while the radius is in units of the instantaneous half-mass
  radius $r_h$. The increase of the probed radial range at large radii
  with time is due to the decrease of $r_h$, while in the central regions
  the different extension is due to the fixed number of stars assumed
  to define the radial bins (see the text). \label{fig1}}
\end{figure}

\begin{figure}[htbp]
\centering
\includegraphics[height=17.0cm, width=17.0cm]{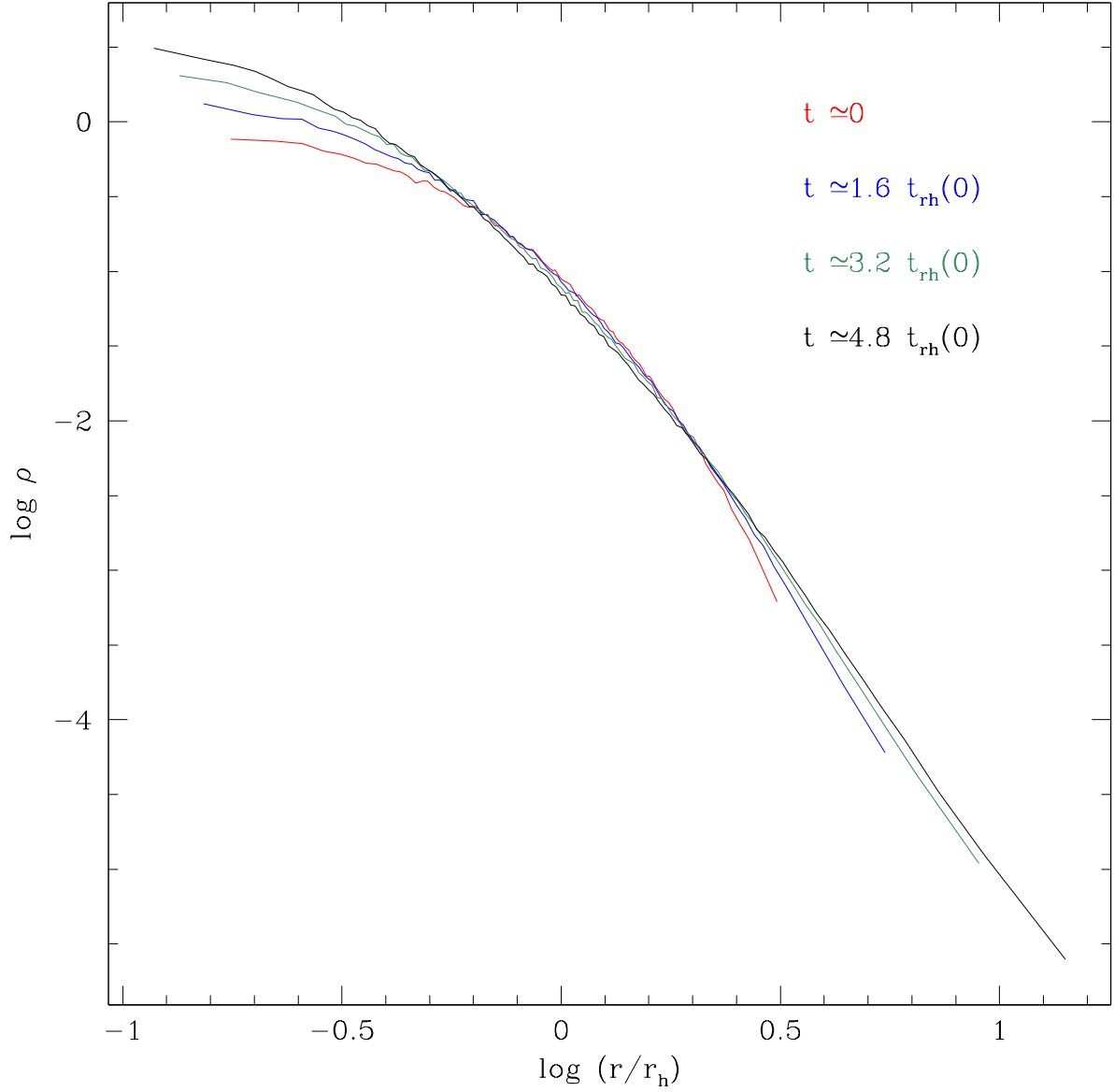}
\caption{Time evolution of the background mass density profile $\rho(r)$ of the
  cluster from the initial time $t = 0$ to the final time $t_f =
  4.8 t_{\rm rh}(0)$, in the mono-mass case derived from the initial
  conditions with $W_0 = 4$. For the color code, and other comments
  see the Caption of \figurename~\ref{fig1}. \label{fig_rho}}
\end{figure}

\begin{figure}[htbp]
\centering
\includegraphics[height=17.0cm,
width=17.0cm]{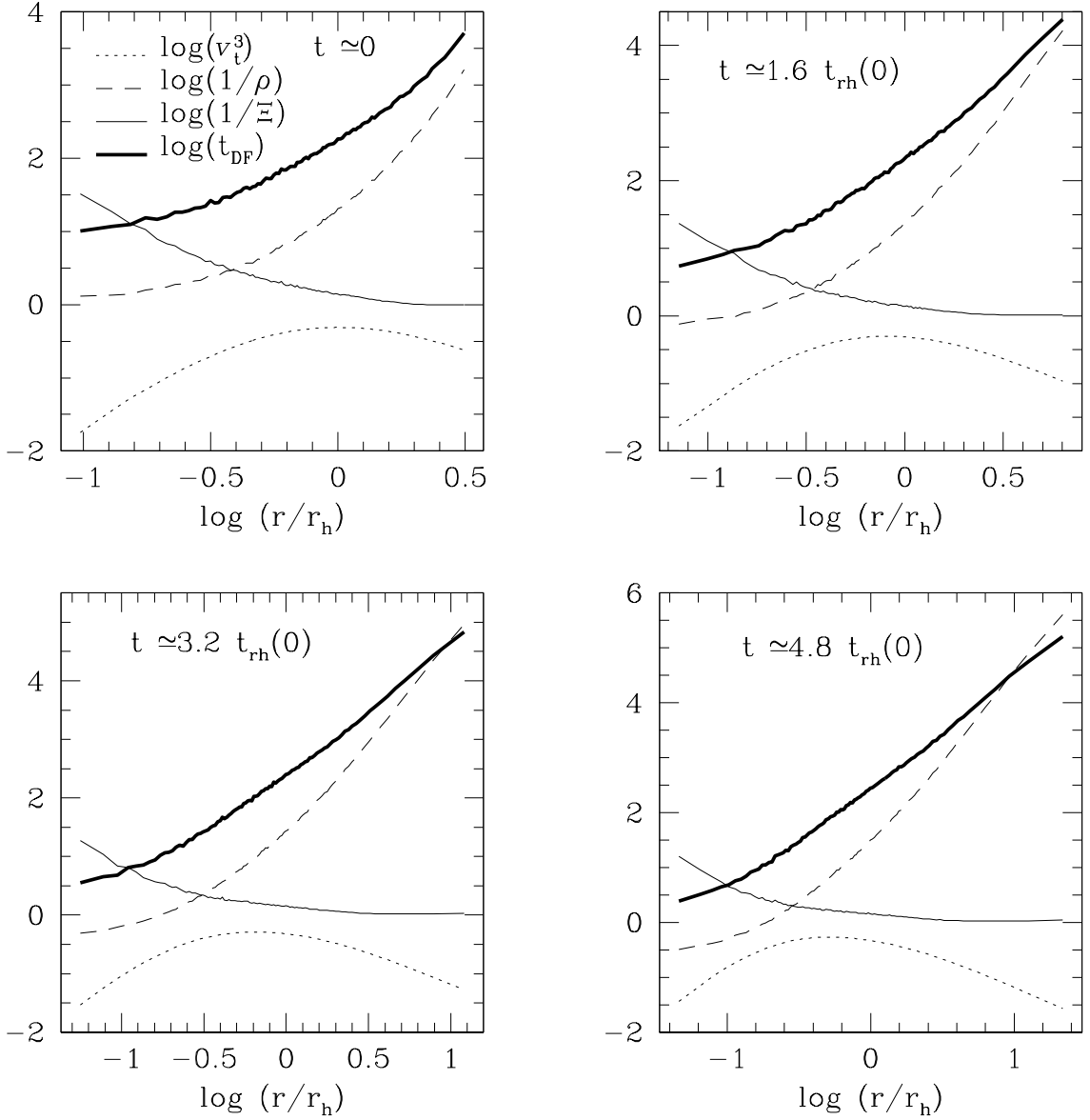}
\caption{Mono-mass case. Radial trend of the DF time-scale (thick
  solid line, eq. \ref{tdf_radiale}) at the four representative times in
  Figures \ref{fig1} and \ref{fig_rho}. In the plots we also show the
  three radially dependent terms contributing to the final value of
  $t_{\rm DF}$ ($v^3_t$: dotted line, $1/\rho$: dashed line, $1/\Xi$: thin
  solid line). All quantities are given here in code units, and from
  eq. (\ref{tdf_radiale}) it follows that $log(t_{\rm DF})$ is just given (modulo an
  additive constant) by the sum of the three quantities.
\label{density_driver}}
\end{figure}

The resulting radial trend of $t_{\rm DF}$ at four representative times is
shown in \figurename~\ref{fig1}, where the radial distance is expressed in
units of the half-mass radius at that time $r_h(t)$, and the time is normalized to
the {\it instantaneous} half-mass relaxation time,
$t_{\rm rh}(t)$. \figurename~\ref{fig1} clearly shows that $t_{\rm DF}$
maintains a monotonic radial trend at increasing time, in agreement with
simple expectations.  We see also that, as the
dynamical evolution of the system proceeds, $t_{\rm DF}$ decreases (i.e.,
DF becomes more efficient) in the central and external regions, while
the effect is opposite at intermediate radii.  This behavior is mainly
due to the time evolution of the mass density (see
\figurename~\ref{fig_rho}), which progressively increases with time in
the innermost and outermost regions, (the former being due to the
core-contraction, the latter due to the related cluster halo
expansion), while it tends to decrease on
the radial interval
$-0.3 \lesssim log (r / r_h) \lesssim 0.3$ (consistently with an
evolution towards a higher cluster concentration, resulting in a
decrease of $r_h$).  Indeed the density
profile is the primary driver of the shape of $t_{\rm DF}(r)$ at all
times, while the other terms entering eq. (\ref{tdf_radiale}) provide
a negligible contribution. This is due to a more significant evolution
of $\rho(r) = m \, n(r)$ with respect to those of $\Xi(r)$ and
$v^3_{t}(r)$, as is clearly apparent in \figurename~\ref{density_driver}.
Therefore, the mono-mass simulations confirm the
scenario adopted by \citet{ferraroNATURE} to explain the different
radial distributions of BSSs as a function of the cluster age, due to the
monotonic increase of the radius at which the cluster age coincides with $t_{\rm DF}(r)$.

\subsection{Multi-mass case}
\label{results-multi}
In the multi-mass case, the DF time-scale has been computed by
evaluating each term in eq. (\ref{somma_armonica}) associated to each
one of the three stellar components considered in the 
simulations described in Sect. \ref{simulazioni_multi}, and by setting
$m_t = 4 m_1$.  To avoid large fluctuations due to the low number
statistics of the most massive component (which counts only $3\%$ of
the total number of particles), the radial binning has been chosen by
imposing that at least $\simeq 200$ stars of mass $m_3$ are included in each
shell and that the total number of bins is always $N_{\rm bin} \ge 19$. Thus,
at odds with the mono-mass case, the total number of particles in each shell
is not constant.  The number density profile of each mass component,
$n_i(r)$, has been computed as in the mono-mass case (see
Sect. \ref{results-mono}), while the circular velocity of the test particle,
is again trivially computed from eq. (\ref{circular}). The
velocity factors $\Xi_i(r)$ in each radial bin have been computed by normalizing the number of stars in
the $i^{th}$ mass group that move slower than $v_t(r)$, to the
total number of particles of the same mass group in that radial bin.

\begin{figure}[htbp]
\centering
\includegraphics[height=17.0cm,width=17.0cm]{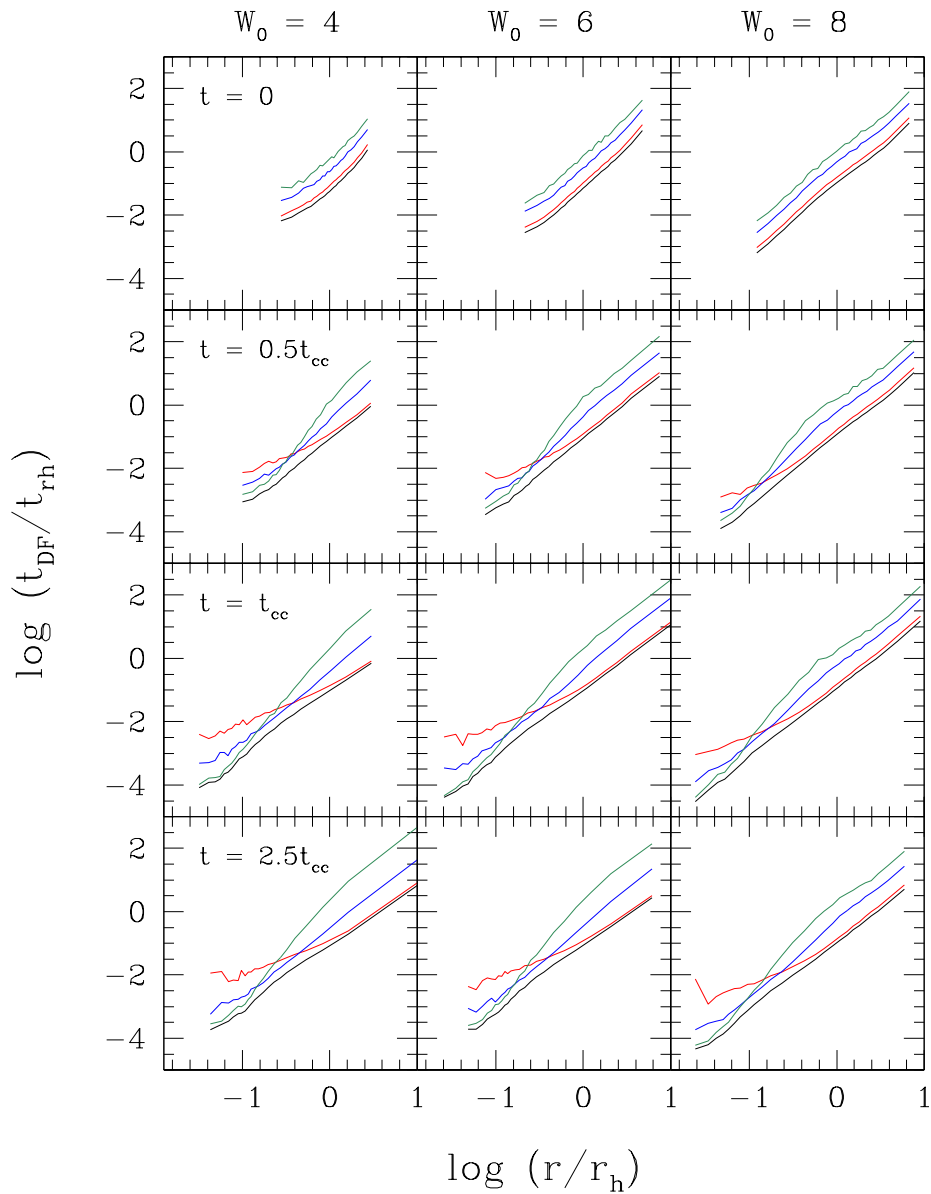}
\caption{Radial profile of $t_{\rm DF}$ in the multi-mass simulations,
  plotted for each mass-component separately (red: $m_1$,
  blue: $m_2$, green: $m_3$) and from the combined effects of the
  three background mass components according to eq. (\ref{somma_armonica}) (black
  lines). The different panels are snapshots taken at $t=0, 0.5, 1$, and
  $2.5 t_{\rm cc}$, where $t_{\rm cc}$ is the fiducial core-collapse time of the system
  (see Section \ref{results-multi}). Panels from left to right, models with: $W_0 = 4$, $6$, $8$. The radial
  distance from the center and $t_{\rm DF}$ are normalized, respectively,
  to the instantaneous half-mass radius and the instantaneous half-mass
  relaxation time computed for the system as a
  whole. \label{single_tdf}}.
\end{figure}

The results obtained for the three considered values of the
dimensionless potential ($W_0 = 4, 6, 8$) and four different times are
plotted in \figurename~\ref{single_tdf}.  In particular, we show
$t_{\rm DF}(r)$ at the initial time of the simulation ($t = 0$) and for
three snapshots around the ``core-collapse time'' $t_{\rm cc}$, defined as
the time at which the Lagrangian radius $r_{\rm 10}$ (i.e., the radius
containing $10 \%$ of the total mass) reaches its minimum value. For
reference, we notice that $t_{\rm cc} \simeq 3.6 t_{\rm rh}(0), 2.5 t_{\rm rh}(0),
0.6 t_{\rm rh}(0)$ for $W_0 = 4,6,8$,
respectively.  The radial behavior of the DF time-scale is plotted
both for each mass-component separately (color lines), and combining
the effects of the three background mass components according to
eq. (\ref{somma_armonica}) (thick black lines). Since the initial conditions
for all mass-groups were built from the same King model properly
scaled only for the adopted number of stars, at $t = 0$ the radial
profile of $t_{\rm DF}$ has the same qualitative behavior for any group
and its value at any radial distance decreases for
decreasing particle mass, because lighter stars are by far the most
numerous and thus dominate the local mass density.

At all evolutionary times, as for the
mono-mass simulations, also in this case the mass density is the main
driver in shaping $t_{\rm DF}(r)$: the local value of $t_{\rm DF}(r)$ is
essentially determined by that of the dominant contributor to
$\rho(r)$. In fact, due to energy equipartition, the most massive
particles progressively migrate toward the center, while the lightest
component expands outward. As a consequence the total mass density is
mainly contributed by the heaviest mass-group at small radii, while it
is dominated by the $m_1$ particles in the outskirts, and the radial
profile of $t_{\rm DF}$ is thus driven by $1/\rho_3(r)$ and $1/\rho_1(r)$
in the two respective radial regions. In addition, from $t = 0$ to $t
= t_{\rm cc}$, DF becomes increasingly more efficient in the center (i.e.,
the value of $t_{\rm DF}$ at fixed small radii becomes smaller) because
also the central mass density increases, due to the segregation first
of $m_3$ particles and then also of $m_2$ stars\footnote{The
  progressive increase of the central density, combined with an
  expansion of the outer layers are also responsible for the enlarged
  radial sampling (both toward the center and in the outskirts) for
  increasing evolutionary times, as a consequence of by the adopted binning
  procedure.}.  The increase of $t_{\rm DF}(r)$ at $t = 2.5 t_{\rm cc}$, is
due to a late re-expansion of the system (corresponding to a decrease
of the mass density).

\figurename~\ref{single_tdf} also shows that the radial profile of
$t_{\rm DF}$ is qualitatively the same for all values of $W_0$ (i.e.,
independently of the cluster concentration) and it always
monotonically increases with radius, as in the mono-mass case. {\it Our
simple analysis then shows that the presence of a mass spectrum does not
induce non-monotonic behaviors in the DF time-scale.}

\subsection{Equivalent Classical System}
\label{equivalent}

\begin{figure}[htbp]
\centering
\includegraphics[height=17.0cm, width=17.0cm]{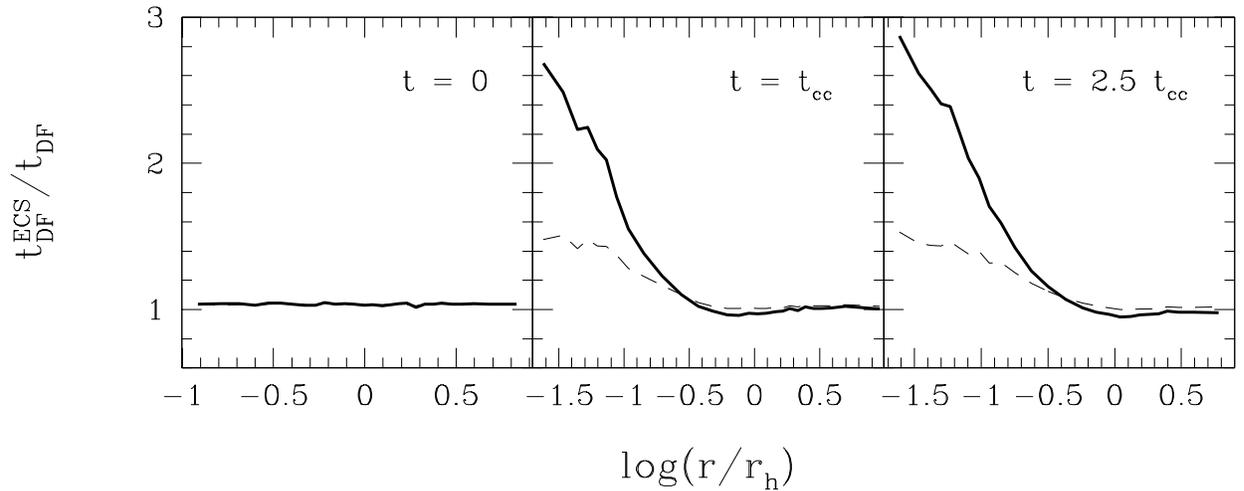}
\caption{Ratio between the DF time-scale computed in the ECS
  approximation and for the multi-mass system, as a function of radius
  (in units of the total half-mass radius), for three evolutionary
  times (see labels) and for the model with initial conditions $W_0 =
  8$. The thick solid lines show the
  results obtained under the assumption that the \emph{number} density
  of the ECS system equals that of the multi-mass case: $n_{\rm ECS}(r) =
  n(r)$. The dashed lines refer to the approximation in which the
  \emph{mass} density of the ECS system is equal to that of the
  multi-mass case: $\rho_{\rm ECS}(r) = \rho(r)$.  The results are
  essentially the same $W_0 = 4$ and $W_0 = 6$.
Notice how the number average estimate can be off by a factor up to
$3$ in the central regions, while a much better estimate is obtained
by using the mass density average. \label{ECS}}
\end{figure}

The previous analysis showed that a mass spectrum in a GC does not
leads to a non-monotonic behavior of $t_{\rm DF}(r)$, so that the results
are in qualitative agreement with the mono-mass case. But what about
the absolute value of $t_{\rm DF}$? In fact, 
Ciotti (\citeyear{ciotti2010}) showed that
a significant overestimate of the DF time-scale can arise if approximating
a multi-mass system with a single component only. In our case this check can
be done by introducing the definition of the {\itshape Equivalent
  Classical System} (ECS):
\begin{itemize}
\item the {\it number} density in the ECS is equal to the {\itshape total} number
          density of the multi-mass case:
\begin{equation}
n_{\rm ECS}(r) = \sum_{i=1}^{N_{\rm pop}} n_i(r);
\end{equation}
\item the mass of the field particles in the ECS ($m_{\rm ECS}$) is equal to the 
{\itshape average} field mass of the multi-component case:
\begin{equation}
m_{\rm ECS}= \frac{1}{N} \sum_{i=1}^{N_{\rm pop}} m_i \, N_i
\end{equation}
where $N$ is the total number of stars in the multi-mass system;
\item the velocity dispersion of the ECS is equal to the 
{\itshape equipartition} velocity of the multi-component case.
\end{itemize}
The last property is non relevant in our case, since we compute the
function $\Xi(r)$ directly from the simulation outputs, as the
fraction of all the stars slower than the test particle.  In order to compare
the DF time-scale obtained in the ECS approximation with the exact
determination for the multi-mass case discussed in the previous
section, we used eq. (\ref{tdf_radiale}) by assuming $n(r) =
n_{\rm ECS}(r)$, $m_t = 4 \ m_1$ and $m = m_{\rm ECS} \simeq 1.2 \ m_1$.  To
compute $M(r)$ entering the definition of $v_t$, we counted the number
of particles with the radius $r$ and multiplied it by $m_{\rm ECS}$.  The
comparison between the DF time-scale in the multi-mass and in the ECS
cases for $W_0 = 8$ is shown in \figurename~\ref{ECS} (solid lines).
As expected, no differences are found at $t = 0$, when mass-segregation has not
played any role yet.  This is due to the fact that the mass density is
nearly the same in the two cases, because the three components have
equal radial distributions and $m_{\rm ECS}$ well corresponds to the true
mean mass at all radii.  When mass-segregation takes place, a
systematic underestimate of DF in the ECS case becomes apparent in the
central regions, where $t_{\rm DF}$ can be almost a factor of $3$ larger
than in the true multi-mass system.  This discrepancy is due to the
systematic underestimate of the ECS local mass density with respect to
the multi-mass one. In fact, as a consequence of mass segregation, the
number density at small radii is mainly contributed by the heaviest
stars, and since $m_3$ is significantly larger than $m_{\rm ECS}$, the ECS
mass density in these regions is considerably smaller than the
multi-mass one. A smaller density implies a lower efficiency of DF,
thus bringing to larger values of $t_{\rm DF}$ in the ECS with respect to
the multi-mass case.  Beyond the half-mass radius, the ratio between
the values of $t_{\rm DF}$ in the two cases becomes nearly flat around
$1$. This is because beyond the half mass radius no appreciable
mass-segregation has taken place during the simulated time, and so the
local average mass is nearly equal to $m_{\rm ECS}$ in that region, as in
the initial conditions.

As a final test we modified the definition of ECS in requiring that
the {\it mass density} in this approximation is equal to the true mass
density of the multi-component system at any radius:
\begin{equation}
\rho_{\rm ECS}(r) = \rho(r) = \sum_{i=1}^{N_{\rm pop}} \rho_i(r).
\end{equation}
As expected the difference between the two representations is
considerably reduced and the underestimate of $t_{\rm DF}$ is about half
that previously found (see the dashed lines in \figurename~\ref{ECS}).
Therefore, we conclude that special attention should be paid to
arguments based on specific requests about the value of $t_{\rm DF}$, as
estimate based on the average properties of the GC can be off by a
factor of $2$ or $3$, factors that are quite important for phenomena
happening on time scales of the order of the Hubble time.

\section{Summary and discussion}
\label{summary}
We studied the radial behavior of the DF time-scale computed by
following the Chandrasekhar's formalism in a background of field
particles with a ``simplified'' mass spectrum, evolved by means of
direct and collisional N-body simulations.  To our knowledge, this is
the first numerical study of DF in the presence of a self-consistently
evolving field of particles with unequal masses and for a test star
with a mass comparable to that of the field (see \citealp{ciotti2010}
for an analytical approach to the problem in the case of a spatially
homogeneous background).  We explored both the mono-mass case (where
all particles have unit mass and the test star decelerated by DF is
four times more massive), and the multi-mass case (with three field
components of masses $m_2 = 2 \, m_1$, $m_3 = 3 m_1$ and total numbers
$N_1$, $N_2$, $N_3$, and with the test particle having $m_t = 4m_1$).
Each simulation was run with $N = 10^4$ particles initially
distributed as a spherical and isotropic King model with dimensionless
potential $W_0=4$ in the mono-mass case, and $W_0=4,6,8$ in the
multi-mass one. To improve the statistics, we combined $20$
simulations differing only for the seed used to generate the initial
conditions.

We find that the radial behavior of $t_{\rm DF}$ is always monotonic
($t_{\rm DF}$ increasing with $r$), both in the mono-mass background
and in the presence of a mass spectrum, independently of the evolutionary
time and the initial concentration of the system. In all cases, the
largely dominant factor in determining the shape of $t_{\rm DF}(r)$ is
the total mass density profile.  

We also find that approximating a multi-mass system as single-mass
cluster (made of stars with masses equal to the average particle mass)
can lead to a systematic overestimate of $t_{\rm DF}$ within the
half-mass radius, up to a factor of 3 in the innermost regions, so
that some care must be used when using the average properties of
background populations to obtain quantitative estimates of $t_{\rm
  DF}$.

The $N-$body simulations and the overall approach presented in this
work are certainly a rough simplification of the much more complex
problem of DF in real GCs. First of all, the total number of simulated
particles is much smaller than the number of stars in a
cluster. However, while this affects the overall cluster evolution
time-scale, it is not expected to impact our conclusions about the
radial monotonicity of $t_{\rm DF}$. The same considerations apply to
the assumptions adopted for the mass spectrum (only three bins with
$m_2= 2 m_1$ and $m_3=3 m_1$): using a larger number of mass groups is
not expected to induce a non-monotonic behavior on $t_{\rm DF}(r)$,
since the density profile $\rho(r)$ would still be a monotonic
decreasing function of radius. Adding a population of black holes,
neutron stars and massive white dwarfs (i.e., stellar remnants more
massive than BSSs, which are certainly present in current GCs) could
make some difference, especially in the cluster central regions where
they are expected to be concentrated because of mass segregation.
However, given the uncertainties on their typical retention fraction,
we preferred to neglect such a component in this first approach to the
problem, postponing this issue to a future paper. In any case, still
because $t_{\rm DF}\propto 1/\rho(r)$, we do not expect that including
massive dark remnants would modify our conclusions substantially.
This also holds for the effect of an external tidal field, which is
mainly expected to favor the evaporation of the lightest stars from
the cluster outskirts.  The impact of a population of primordial
binaries and multiple systems, instead, is more difficult to predict
(especially if also BSSs are modeled as binary systems) and we plan to
explore this problem in a forthcoming paper.

At least within the adopted approximations, the monotonic behavior
found in all cases for $t_{\rm DF}(r)$ appears to be a quite solid
result. We therefore conclude that, at present, the scenario proposed
by \citet{ferraroNATURE}, where $t_{\rm DF}(r)$ is implicitly assumed
to be monotonic at all times and DF progressively affects larger
clustercentric distances as a function of time, still appears to be
the most likely explanation of the observed BSS radial distributions,
further confirming that these are a powerful empirical tool able to
measure the dynamical age of stellar systems.

\section*{Aknowledgements}
This research is part of the project Cosmic-Lab (web site:
http://www.cosmic-lab.eu) funded by the European Research Council
(under contract ERC-2010-AdG-267675). L.C. was supported by PRIN MIUR
2010-2011, project "The Chemical and Dynamical Evolution of the Milky
Way and Local Group Galaxies", prot. 2010LY5N2T. We thank the
anonymous Referee for useful comments that improved the presentation
of the paper.

\newpage

\end{document}